\documentclass{aa} 
\usepackage{caption}
\usepackage{amssymb}
\usepackage{enumerate}
\usepackage{subfigure}
\usepackage{booktabs}
\bibpunct{(}{)}{;}{a}{}{,}

\usepackage{natbib,twoopt}
\usepackage[breaklinks=true]{hyperref} 

\def\apj{ApJ}%
\def\apjl{ApJ}%
\def\apjs{ApJS}%
\def\ao{Appl.~Opt.}%
\def\aap{A\&A}%
\def\icarus{Icarus}%
\def\mnras{MNRAS}%
\def\pasp{PASP}%
\def\gca{Geochim.~Cosmochim.~Acta}%
\begin{document}

   \title{C/O vs. Mg/Si ratios in solar type stars: The HARPS sample}

   \author{L. Su\'arez-Andr\'es 
          \inst{1,2,3}\fnmsep\thanks{\textit{Send offprint requests to}: L. Su\'arez-Andr\'es,
                \newline \email{lsuarez@ing.iac.es}}, G. Israelian\inst{1,2}, J.I. Gonz\'alez Hern\'andez \inst{1,2}, V. Zh. Adibekyan \inst{4},\newline E. Delgado Mena \inst{4}
                        , N. C. Santos \inst{4,5}
                \and
                        , S. G. Sousa \inst{4,5}
          }

   \institute{Instituto de Astrof\'isica de Canarias, E-38205 La Laguna,
Tenerife, Spain\\
              \and
           Depto. Astrof\'isica, Universidad de La Laguna (ULL),
E-38206 La Laguna, Tenerife, Spain\\
\and
Isaac Newton Group of Telescopes, Apartado de Correos 321, E-38700 Santa Cruz de la Palma, Spain\\
\and
          Instituto de Astrof\'isica e Ci\^encias do Espa\c{c}o, Universidade do Porto, CAUP, Rua das Estrelas, 4150-762 Porto, Portugal \\
         \and 
         Departamento de F\'isica e Astronomia, Faculdade de C\^iencias, Universidade do Porto, 4169-007 Porto, Portugal} 

   \date{Received March 7, 2017; accepted January  14, 2018}

 
  \abstract
   {}
   {We aim to present a detailed study of the magnesium-to-silicon and carbon-to-oxygen ratios (Mg/Si and C/O) and their importance in determining the mineralogy of planetary companions.}
   {Using {499} solar-like stars from the HARPS sample, {we determined C/O and Mg/Si elemental abundance ratios to study the nature of the possible planets formed. We separated the planetary population in low-mass planets ( < 30 $\rm M_{\odot}$) and high-mass planets ( > 30 $\rm M_{\odot}$) to test for a possible relation with the mass.}}
   {We {find} a diversity of mineralogical ratios that reveal the different kinds of planetary systems that can be formed, most of them dissimilar to our solar system. The different values of the Mg/Si and C/O can determine {different} composition of planets formed. {We found that 100\% of our planetary sample present C/O < 0.8. 86\% of stars with high-mass companions present 0.8 > C/O > 0.4, while 14\% present C/O values lower than 0.4. {Regarding Mg/Si, all stars with low-mass planetary companion showed values between one and two, while 85\% of the high-mass companion sample does. The other 15\% showed Mg/Si values below one.  No stars {with planets} were found with Mg/Si > 2}. Planet hosts with low-mass companions present C/O and Mg/Si similar to those found in the Sun, whereas stars with high-mass companions have lower C/O.}}
  {}
  

\titlerunning{C/O vs.\\  Mg/Si ratios in solar type stars: The HARPS sample.}
\authorrunning{Su\'arez-Andr\'es, L. et al.}
\maketitle
%
\section{Introduction}

The determination of the chemical composition of extrasolar planets has been the subject of numerous studies in recent years. One of the keystones has been the fact that {stars hosting giant planets} are considered to be metal-rich when compared with single stars \citep{gonzalez97,santos01,santos04,fischer05,gonzalez06}. Stars hosting low mass planets (those with masses below ~30 $\rm M_{\odot}$) do not seem to be preferentially metal-rich \citep[e.g.][]{sousa08,sousa11a}

{Recent studies of chemical abundances in stars with and without planets have shown no important differences in [X/Fe] vs. [Fe/H] trends between the two groups of stars for refractory \citep{jonay10,elisa} and volatile elements \citep{nitrogeno, carbono}. On the other hand, studies by \citet{haywood08, haywood09} and \citet{adi_over} found evidence of $\alpha$-elements enhancement in stars with planets.} 

As both planetesimals and planets are formed within {the same environment}, their composition is expected to be the same as that of their host star. This assumption might be true for refractory species, but not for the volatile ones \citep{lodders03,thiabaud14}. Although this fact does not affect the magnesium-to-silicon ratio (Mg/Si) for rocky planets, it can affect the carbon-to-oxygen ratio (C/OP \citep{dorn15,thiabaud15b}. {Elemental abundance ratios (C/O and Mg/Si) are important as they govern the distribution and formation of chemical species in the protoplanetary disc.}

Variations in C/O can be found among planetary atmospheres and host stars \citep{mad12, konopacky13, moses13, brewer16} and among planets in the same planetary system. This is due to different parameters (temperature, pressure, etc.) and processes at work during the planet formation stage, including possible migrations of planets from their birthplace \citep{oberg, alidib14, mad12, thiabaud15a,thiabaud15b, brewer16}

In the last few years, several studies have tried to understand the formation and evolution of planets using theoretical models. \citet{bond10, bond10b}, \citet{carterbond} {studied} planet formation scenarios with different initial composition, {but they did not carry out a detailed study of the output volatile species.} \cite{elser12} tried to study this planet formation {model by} adding the formation of solids, but they were unable to reproduce some features  present in the solar system, such as high Fe on Mercury.

Recently, \citet{thiabaud14, marboeuf14} made a complete study, presenting models not only for refractory species, but for volatiles as well.  \cite{thiabaud15a} presented a complete study, taking into account the accretion of several compounds omitted in previous works such as He, $\rm H_{2}$,  $\rm H_{2}O$,  $\rm CO$, $\rm CO_{2}$, $\rm CH_{3}OH$, $\rm CH_{4}$, $\rm NH_{3}$, $\rm N_{2}$, and $\rm H_{2}S$. Also, they computed the importance of volatile elements in protoplanetary discs and their implications in planetary formation \citep{thiabaud15b}. {Their models show that the condensation of volatile species as a function of radial distance allows for C/O enrichment in
specific parts of the protoplanetary disc of up to four times the solar value. This could lead to the formation of planets that can be enriched in C/O in their envelope up to three times the solar value. Their models are consistent with recent observations of hot-Jupiter atmospheres \citep{brewer16}}.

The amount of carbides and silicates formed in planets is controlled by the carbon-oxygen ratio {\citep[e.g.][]{bond10}}

\begin{itemize}
        \item if C/O $< 0.8$: Si will form solid  $\rm SiO_{4}^{4-}$ and $\rm SiO_{2}$, {serving as seeds for}
        Mg silicates {for which the exact composition will be determined by Mg/Si.}
        \item if C/O $> 0.8$: Si will be solid as SiC. Also, graphite and TiC will be formed.
        
\end{itemize}

Silicates are an important ingredient in the formation of rocky planets, as they are the most abundant {compounds} in the mantle and crust of these planets \citep{morgan80}. Silicate distribution is ruled by Mg/Si of the planet host star. Concerning Mg/Si, the principal components could be, as proposed by \citet{bond10b}:
\begin{itemize}
        \item if Mg/Si $<$ 1, Mg forms orthopyroxene ($\rm MgSiO_{3}$) and the remaining Si forms other minerals, such as feldspar ($\rm CaAl_{2}Si_{2}O_{8}$, $\rm NaAlSiO_{8}$) or olivine($\rm Mg_{2}SiO_{4}$)
        \item if $1<$ Mg/Si $<$ 2,  Mg is distributed equally between pyroxene and olivine.
        \item if Mg/Si $>$ 2, Si forms olivine and the remaining Mg forms other oxides like $\rm MgO$.
\end{itemize}
 Testing and improving planetary formation models is key to future studies of habitability, as these ratios are essential elements in defining the structure of the planet.
 
In this article we present a study of C/O and Mg/Si in solar-type stars and their {implications for possible terrestrial planetary formation.}

\section{Sample description}

The high-resolution spectra analysed in this article were obtained with the HARPS spectrograph at La Silla Observatory (ESO, Chile) during the HARPS GTO, HARPS-2 and HARPS-4 programmes \citep[see][for more information about the programme]{mayor03, locurto10, santos11}. The spectra have been used previously in the analysis of stellar parameters, as well as the derivation of precise chemical abundances \citep[see e.g.][]{sousa08,sousa11a, sousa11b, adi12b, tsantaki,sara, carbono}.

{The studied sample is part of the 1111-star sample presented by \citet{adi12b} for which the abundance of volatiles were measured \citep{sara,carbono}. We limited our metallicity sample for stars with [Fe/H] > -0.6, which is the lower [Fe/H] for the planet-host sample. We studied {499} FGK solar-type stars, with effective temperatures between 5250 K and 6666 K, metallicities from $-0.59$ to $0.55$ dex, and surface gravities from $3.81$ to $4.82$ dex. } 

Of {499} stars, 99 are planet hosts\footnote{Data from www.exoplanet.eu {and the SWEET-CAT catalogue (www.astro.up.pt/resources/sweet-cat/)}}, whereas the other {400} are single stars (stars with no known planetary companion, {also known as comparison stars}. {We urge the reader to take into account that stars considered as single stars might harbour undetected planets \citep[see][]{mayor11,faria16}}. 

\section{Chemical abundances}

To obtain the carbon-to-oxygen (C/O) and magnesium-to-silicon (Mg/Si) elemental abundance ratios we adopted chemical abundances from \cite{adi12b, sara} and \cite{carbono} for Mg and Si, O, and C, respectively. All these chemical abundances were obtained using the same high-resolution high-quality HARPS spectra {and the same stellar parameters}, allowing us to obtain precise elemental abundance ratios. For carbon we studied the CH molecular band located at 4300 \AA \space \citep[see][]{carbono}, whereas for the other elements atomic features were used. {Recent studies of molecular CH features \citep{carbono} have proved that they are as reliable as atomic ones, and provide consistent results.}

{Although the adopted abundances come from different sources, O, Mg, and Si abundances were obtained following the same procedure (EW method). For the case of C, synthetic fitting was applied, but compared to the EW method results to confirm their validity \citep[see][]{carbono}. Errors are significantly lower for Mg and Si (averages 0.06 and 0.04, respectively) than for C and O (average 0.15 for both cases), which affects our final elemental abundance ratio errors. }

{Adopted solar abundances for all elements are $\log\epsilon(\rm C)_{\odot}=8.50$ dex \citep{caffau}, $\log\epsilon(\rm O)_{\odot}=8.71$ \citep{caffau08}, $\log\epsilon(\rm Mg)_{\odot}=7.58$ dex and $\log\epsilon(\rm Si)_{\odot}=7.55$ \citep{anders89}.}

        \section{Elemental abundance ratios}
        
        {Theoretical studies suggest that C/O and
                Mg/Si are very important in determining the mineralogy of terrestrial planets. {Since the mineralogy of planets is commonly studied in terms of absolute ratios, we use these instead of solar ratios\footnote{We remind the reader that C/O $\neq$ [C/O]. Please see Section 6 for more information.}.}}

        Elemental abundance ratios were calculated using the following equation

                \begin{equation}
        A/B=N_{A}/N_{B}=10^{log\epsilon(A)}/10^{log\epsilon(B)}
        ,\end{equation}

        where  log$\epsilon(A)$ and log$\epsilon(B)$ are absolute abundances. Errors were estimated by evaluating an increase or decrease in the $log\epsilon(A)-log\epsilon(B)$ abundance ratio, due to the relative errors (For more details, see \citealt{elisa}). {Median errors are represented in all figures.(0.14 for C/O and 0.06 for Mg/Si))} A sample table
        is shown in Table \ref{res}.

                \begin{table}[!h]\scriptsize
                
                \caption{Sample of C/O and Mg/Si abundances for a set of stars (see online table).}
                
                \begin{center}
                        \scalebox{0.9}[1]{
                                \begin{tabular}{l c c  c c c}
                                        \hline
                                        \hline
                                        
                                        Star    &       $\rm T_{\rm eff}$       &       log $g $& [Fe/H]& C/O & Mg/Si \\
                                        \midrule

                                        HD39091&    5991&4.40&0.09&$0.61\pm {0.11}$& $1.11\pm {0.10} $\\
                                        
                                        HD82943 &    5992&4.42&0.28&$0.63\pm { 0.13}$& $0.99\pm {0.04}$ \\
                                        HD100508&    5384&4.39&0.35&$0.54\pm { 0.09}$& $0.91\pm {0.01}$ \\
                                        HD100777&    5530&4.31&0.25&$0.53\pm { 0.04}$& $0.96\pm {0.12}$\\

                                        HD102117&    5620&4.28&0.26&$0.48\pm { 0.09}$& $1.07\pm {0.01}$ \\
                                        HD212708&    5644&4.31&0.26&$0.51\pm {0.09}$& $1.06\pm { 0.07} $\\

                                        \hline
                        \end{tabular}}
                        \label{res}
                \end{center}
        \end{table}
        
        \subsection{C/O}
        
        {The C/O in planet-host stars can provide key information about the protoplanetary disc in which the planet was formed. The dependence on} the distance for volatile elements will affect the C/O  expected in exoplanetary atmospheres, as volatiles are heavily affected by different distances (or different ice line positions) during the early lifetime of the nebula while planets accrete \citep{oberg, alidib14, brewer16}.

        Figure \ref{co} shows the C/Os derived in this paper as a function of [Fe/H] for {these} samples; stars with and without planets. We obtain a linear fit for both samples, with very little differences between them. {We can see a dependence of C/O on metallicity, in agreement with other works, such as \cite{nissen14}, \cite{teske14} and \cite{brewer}.}
        
        {Several authors have studied }the C/O ratios in stars with planets \citep[e.g.][]{petigura, nissen13, nissen14, teske14}. In Fig. \ref{nissen} we can see our results {compared} with the work of \cite{nissen14}, \cite{teske14}, and \cite{jonay13}. {To make this comparison possible, we scaled the carbon and oxygen abundances presented by these authors to our reference values (log$N_{\odot}$(O) = 8.71 and log$N_{\odot}$(C) = 8.50). {As seen in Fig. \ref{nissen}, there is a good fit at all metallicities between our work and these previous studies.}              \begin{figure}[!h]
                        \begin{center}
                                \includegraphics[angle=0,width=1.05\linewidth]{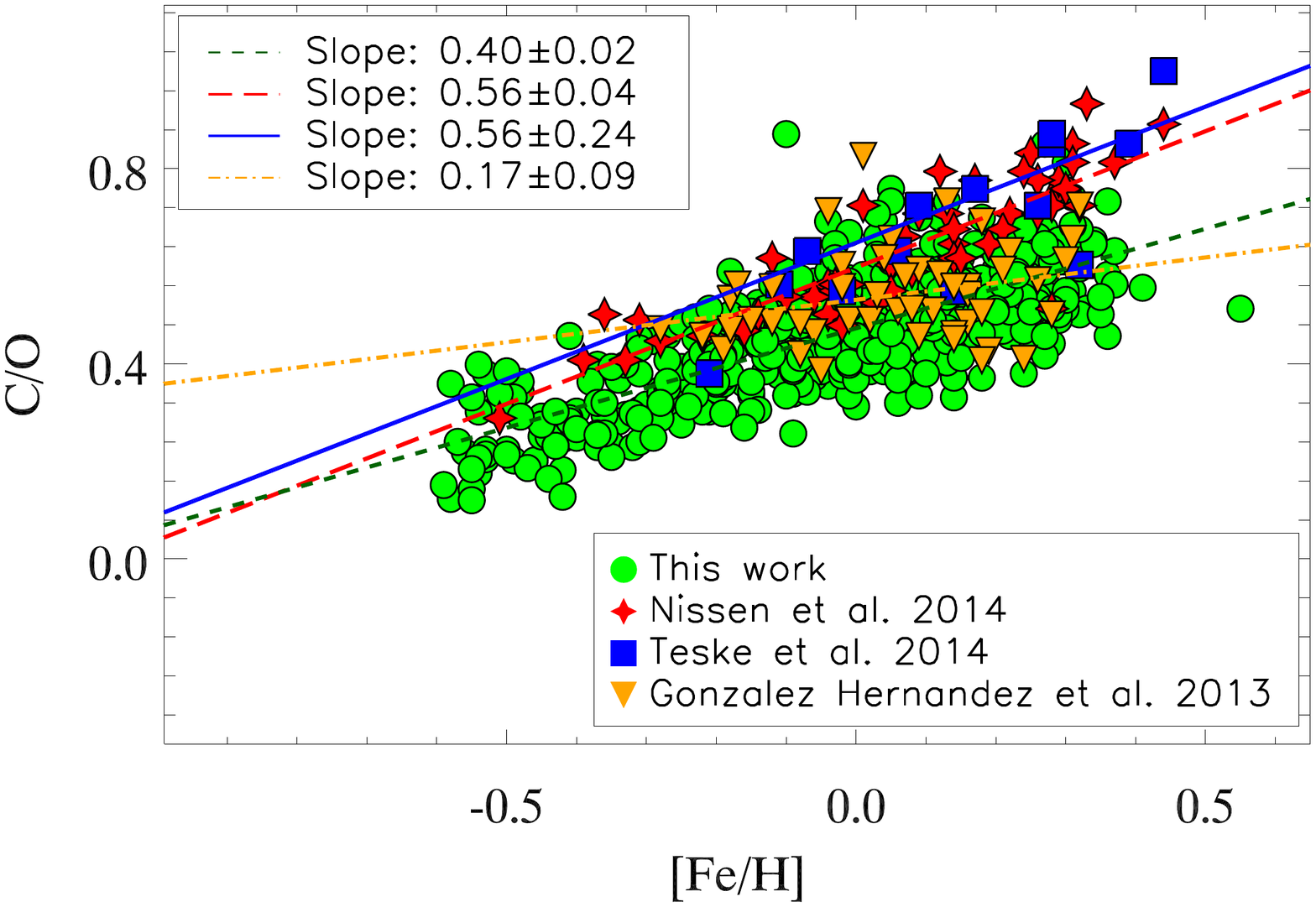}
                                        \caption{Comparative C/O versus [Fe/H]. Green dots show data from our work; red stars show results from \cite{nissen14}; blue squares from \cite{teske14}; orange triangles  from \cite{jonay13}}

                                \label{nissen}
                        \end{center}
                \end{figure}

        \begin{figure}
                \begin{center}
                        \centering
                        \scalebox{1.15}[1.15]{
                                \includegraphics[width=80mm, angle=180]{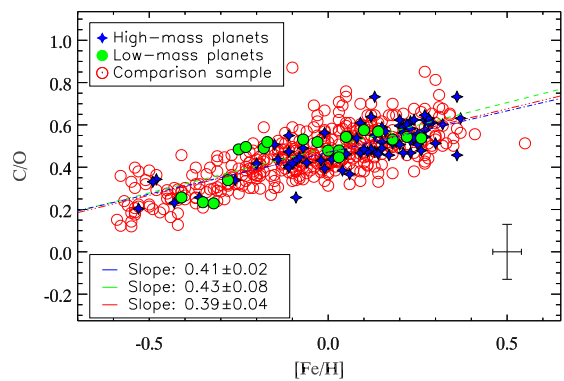}}
                        \caption{C/O versus [Fe/H]. Red open circles refer to single stars while green dots refer to low-mass planet host stars and blue diamonds to high-mass planet-hosts.}
                        \label{co}
                \end{center}
        \end{figure}
        
        {To test for possible relationships between C/O and the masses of planetary companions, 
                we separated the planet population into two groups: low-mass planets (LMP; with masses less than or equal
                to 30 $M_{\oplus}$) and high-mass planets (HMP; with masses greater than 30 $M_{\oplus}$). In those stars which host several planets, the most massive planet in the system was considered in our study. Our sample consists of 19 low-mass and 80 high-mass planet hosts.}
        If we take a closer look at the C/O distribution for the studied samples in Fig. \ref{histo_co}, we can see that the planetary samples exhibit almost the same behaviour, with  an small offset, while the single star sample spreads more, with a significant higher FWHM (see Table \ref{stats}). In our sample, $100\%$ of stars with planets have C/O values lower than 0.8. {Average value for the low-mass and high-mass samples are 0.46$\pm$0.11 and 0.50$\pm$0.10, respectively. Average value for the single stars sample is 0.45$\pm$0.13. Our stars are slightly carbon poor when compared to the solar reference ($\rm C/O_{\odot}$=0.61). These results agree, within errors, with those presented in \cite{nissen14, teske14,jonay13}. }

        \begin{figure}
                \begin{center}
                        \scalebox{1.1}[1.1]{
                                \includegraphics[width=80mm, angle=180]{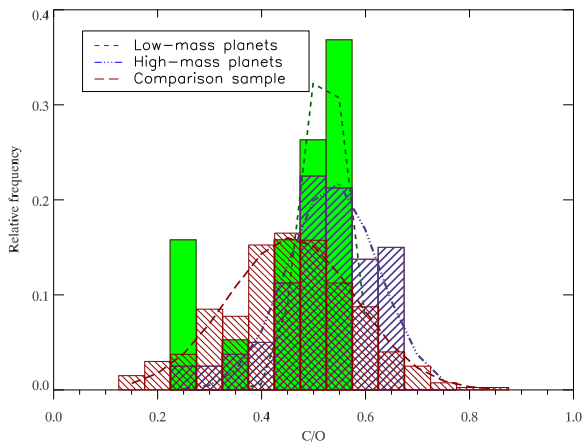}}
                        
                        \caption{C/O distributions for stars harbouring low-mass (green) and high-mass (blue) planets. Stars without planets are shown in red. }
                        \label{histo_co}
                \end{center}
        \end{figure}
                     \begin{figure}
                        \begin{center}
                                \scalebox{1}[1]{
                                        {\includegraphics[width=80mm,angle=180]{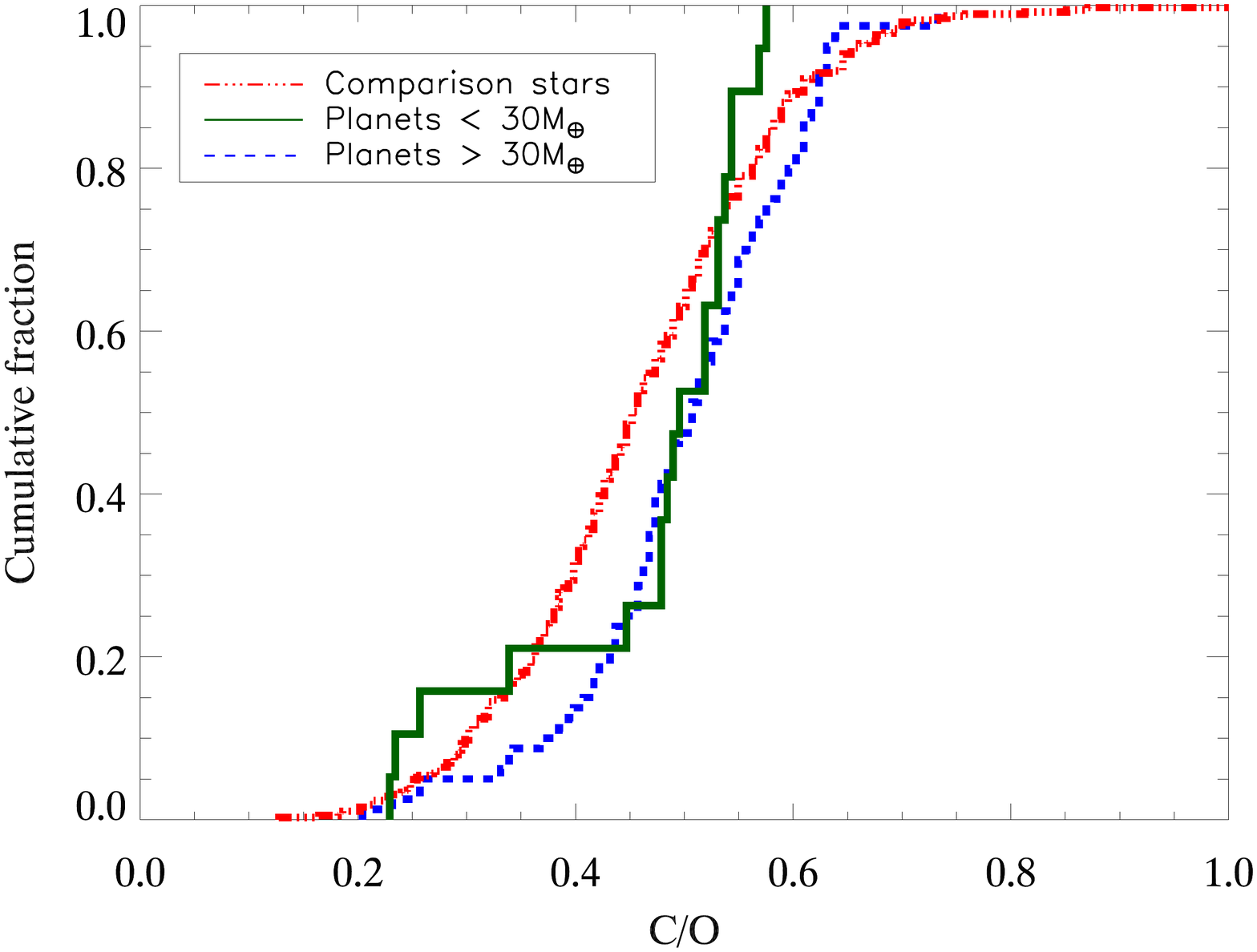}}}
                                \caption{Cumulative C/O distributions for single stars (red) and stars with planets (low-mass (green) and high-mass planets (blue)).}
                                \label{3sample}
                        \end{center}
                \end{figure}

                                \begin{table}\small
                        
                        \caption{Statistics for the fitted histograms presented in Figures \ref{histo_corr}, \ref{histo_co}, and \ref{mgsi_histo}. The low-mass planets 
                                (LMP) and the high-mass planets sample (HMP)}
  
                        \begin{center}
                                \scalebox{1.}[1.]{
                                        \begin{tabular}{l l c c}
                                                \hline
                                                \hline
                                                
                                                &Sample & Centre-fit & FWHM \\
                                                \midrule
                                                C/O & Single stars & 0.45 & 0.29 \\
                                                &   Planets: LMP & 0.52 & 0.10 \\
                                                &   Planets: HMP & 0.52 & 0.21 \\
                                                Mg/Si & Single stars & 1.11 & 0.22 \\
                                                &   Planets: LMP & 1.06  & 0.25 \\
                                                &   Planets: HMP & 1.08 & 0.16 \\
                                                
                                                $\rm [C/O]_{\rm corr}$ & Single stars & -0.11 & 0.25 \\
                                                &   Planets: LMP & -0.01 & 0.11 \\
                                                &   Planets: HMP & -0.03 & 0.17 \\
                                                
                                                \bottomrule
                                \end{tabular}}
                                \label{stats}
                        \end{center}
                \end{table}
                
        We can see in Fig. \ref{histo_co} the C/O distributions for single stars and both planetary samples. A cumulative histogram (Fig. \ref{3sample}) shows us that each sample behaves in a different way. We performed a Kolmogorov -Smirnov test (K--S test) to confirm these  behaviours. The null hypothesis is rejected at level $\alpha$=0.1 if our results are higher than $D_{nn'}$=0.29 and $D_{nn'}$=0.15, for the low-mass and high-mass samples, respectively, following the expression
                \begin{equation}
                D_{nn'} > c(\alpha) \sqrt{\frac{n+n'}{nn'}} ; c(\alpha = 0.10) = 1.22
                .\end{equation} 
                {We can see in Table \ref{table:ks_co} that the C/O is above the threshold limit. The probabilities of similarity are 4\% and 0\% for the low and high-mass samples, respectively, so we can assume that the samples do not come from the same distribution.}
        \begin{table}
                        
                        \caption{K--S test for all the samples.}

                        \begin{center}
                                \scalebox{1.}[1.]{
                                        \begin{tabular}{l c c c}
                                                \hline
                                                \hline
                                                
                                                Sample & C/O & Mg/Si\\
                                                \midrule
                                                LMP - NP &     0.31  &    0.16 \\
                                                HMP - NP &  0.25 &    0.12  \\

                                                \bottomrule
                                \end{tabular}}
                                \label{table:ks_co}
                        \end{center}
                \end{table}

                \subsection{Mg/Si}
                
                The magnesium and silicon elemental abundance ratio controls the exact composition of silicates  that can be found in the planetary companion, as this ratio, along with Fe/Si, does not depend so strongly on the distance to the star as the C/O ratio does \citep{thiabaud15a, thiabaud15b, dorn15}.

                In Fig. \ref{mgsi} top panel we show Mg/Si ratios as a function of metallicity. As expected, the Mg/Si ratio decreases with [Fe/H]{, with a similar slope for all the studied sample (see Table \ref{table:stats1}}) We also {find} a relation between values of Mg/Si and $\rm T_{\rm eff}$, as shown in the bottom panel of Fig. \ref{mgsi}. 
                {We obtained slopes and standard deviations for the three studied samples (see Table \ref{table:stats1}) and, due to the high scatter of the single star sample at $\rm T_{eff}$ higher than 6100K, we cannot conclude any physical difference. However, as we can see that that slopes and scatter increase, we decided not to use these stars with $\rm T_{eff}$ higher than 6100K to avoid introducing errors to our analysis.}

\begin{figure}
                        \centering
                        \includegraphics[width=1.05\linewidth, angle=180]{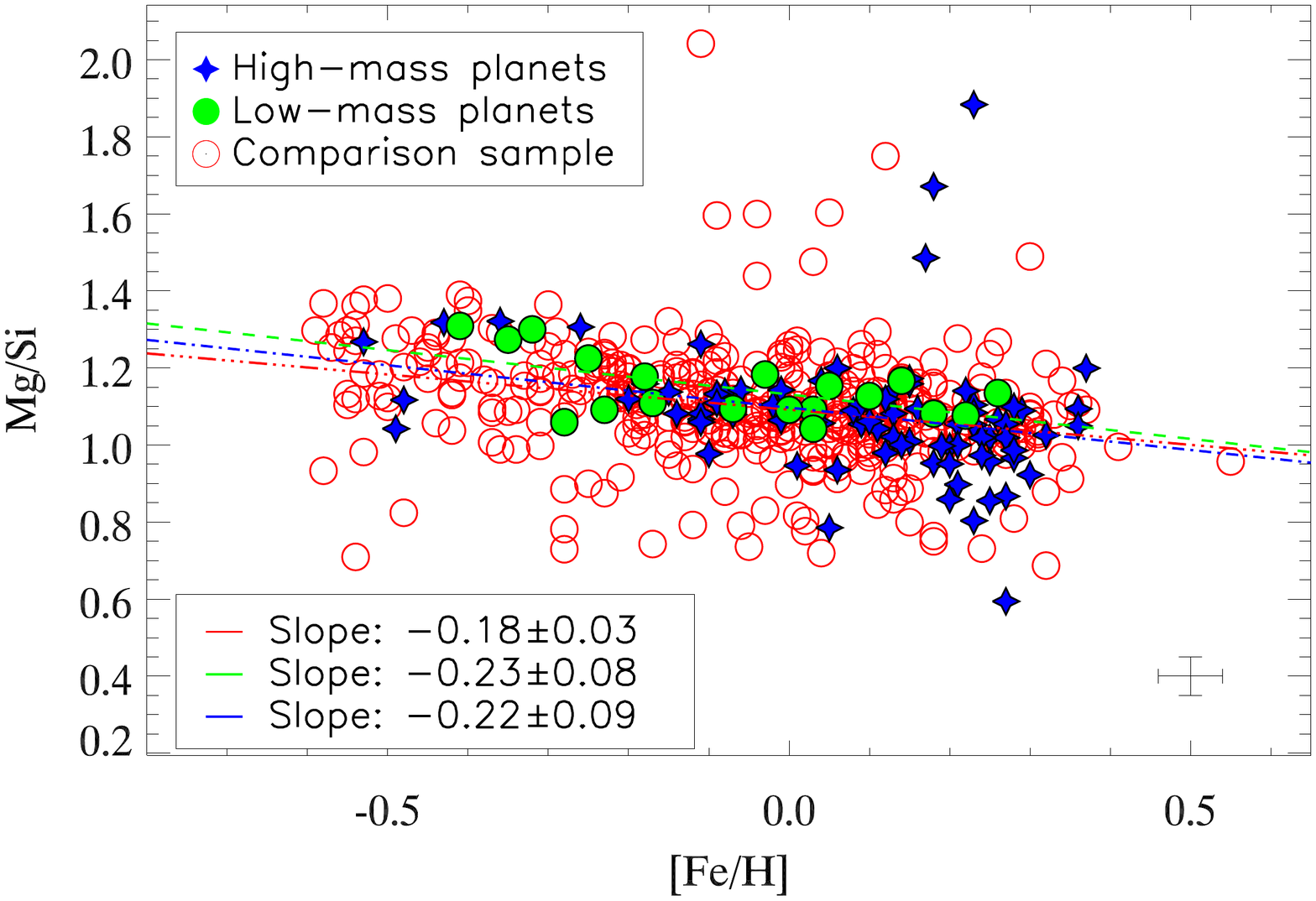}
                        \includegraphics[angle=180,width=1.05\linewidth]{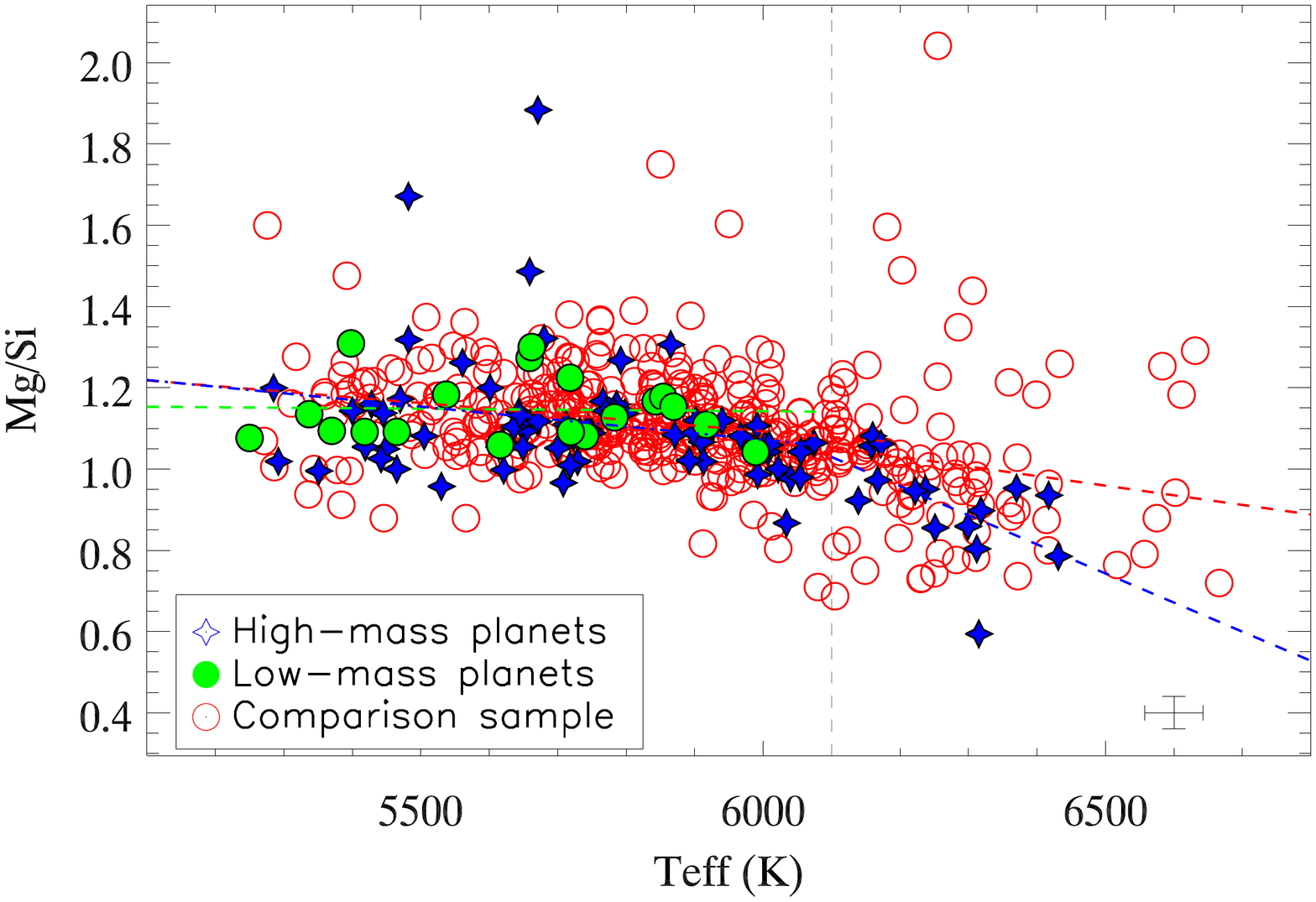}
                        \caption{Mg/Si as a function of [Fe/H] (top panel) and $\rm T_{\rm eff}$ (bottom panel). Red open circles refer to single stars while green dots refer to low-mass planet host stars and blue diamonds to high-mass planet-hosts. }
                        \label{mgsi}
                        
                \end{figure}

                {Given the uncertainty of the implication of NLTE corrections for these hot stars, we will study, from now on, only the sample with $\rm T_{eff} < 6100$K.} {The average value for the low-mass and high-mass samples are 1.15$\pm$0.08 and 1.12$\pm$0.15, respectively. The average value for the single stars sample is 1.12$\pm$0.12. These results agree, within errors, with those presented in \cite{brewer16}}.

                        \begin{table}[!t]\small
                        \caption{Statistics for the three groups studied (Comparison sample, 
                                low-mass planets and high-mass planets) for Mg/Si vs. $\rm T_{\rm eff}$ (see Fig. \ref{mgsi}, bottom panel).}             
                        \label{table:stats1}      
                        \centering   
                        \scalebox{1.1}[1.1]{      
                                \begin{tabular}{  c c c } 
                                        \hline
                                        \hline 
                                        
                                        & \multicolumn{1}{c}{Mg/Si vs. $\rm T_{\rm eff}$}\\
                                        & \multicolumn{1}{c}{{\scriptsize {(for $\rm T_{\rm eff}$ < 6100 )}}}\\
                                        \cmidrule{2-3} 
                                        &   Slope & Slope$\rm_{\rm err}$ \\
                                        
                                        \cmidrule{2-3}        
                                        NP  & -1.39E-04 & 0.32E-04 \\
                                        LMP  & -1.21E-05 & 8.98E-05\\
                                        HMP  & -1.63E-04& 0.90E-04 \\    
                                        \cmidrule{1-3}                    
                                        &  \multicolumn{1}{c}{Mg/Si vs. $\rm T_{\rm eff}$}&\\
                                        &  \multicolumn{1}{c}{{\scriptsize {(for $\rm T_{\rm eff}$ > 6100 )}}} &\\
                                        \cmidrule{2-3} 
                                        &   Slope & Slope$\rm_{\rm err}$  \\
                                        \cmidrule{2-3}        
                                        NP  & -2.37E-04 & 2.54E-04  \\
                                        LMP  &  -& -  \\
                                        HMP  & -7.17E-04 & 2.97E-04 \\    
                                        \cmidrule{1-3}                    
                                        
                                        \hline                
                        \end{tabular}}

                \end{table}
                
                \begin{figure}[!h]
                        \begin{center}
                                \scalebox{1.}[1.]{
                                        {\includegraphics[width=80mm, angle=0]{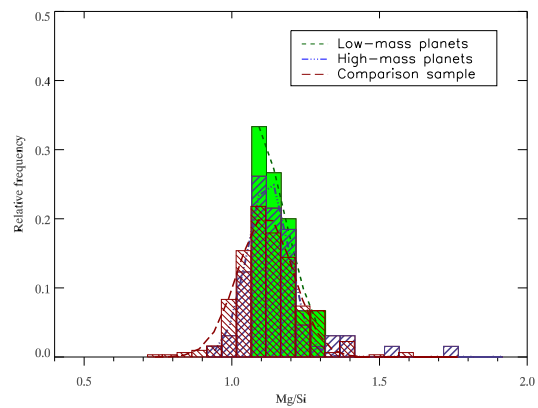}}}
                                \caption{Mg/Si distributions for stars harbouring low-mass (green) and high-mass (blue) planets. Stars without planets are shown in red. }
                                \label{mgsi_histo}
                        \end{center}
                \end{figure}

        {       {The distribution of Mg/Si for both planet-host stars and the comparison sample are shown in Figure \ref{mgsi_histo}. We can see that {all samples} exhibit the same behaviour. In our sample 100\% of the low-mass sample have an Mg/Si value of between 1.0 and 2.0 while 85\% of the high-mass sample exhibit this behaviour. We also find that none of the low-mass planet hosts have Mg/Si values below 1.0, but 15\% of the high-mass sample does. }
                        No stars with Mg/Si values greater than 2.0 were found. }
                
                {Figure \ref{mgsi3sample} shows Mg/Si distributions for single stars and both planetary samples. All samples are concentrated around Mg/Si$\sim$1.12, only 0.05 higher than our solar reference (1.07) and within errors. Statistics for these histograms are shown in Table \ref{stats}. A cumulative histogram (Fig. \ref{mgsi3sample}) shows that each sample behaves in the same way. We find no differences between the planetary samples, as all three samples peak around 1.10. The high-mass sample has a broader distribution than the low-mass one, but the limited number of the low-mass sample does not allow us to infer any result other than the similarity between the samples. A similar result can be found in \cite{brewer16}.}

                \begin{figure}
                        \begin{center}
                                \scalebox{1}[1]{
                                        {\includegraphics[width=80mm,angle=180]{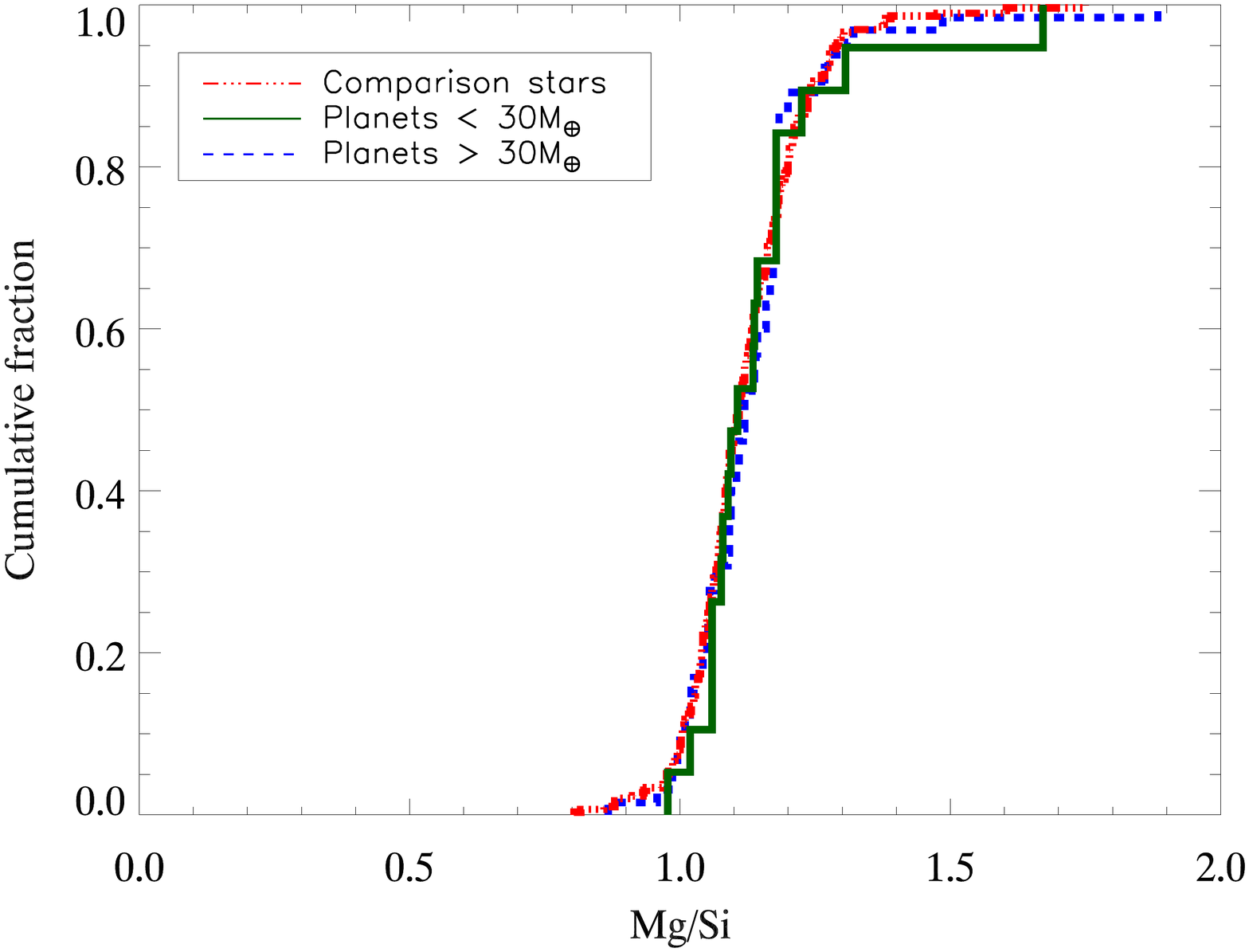}}}
                                \caption{Cumulative Mg/Si distributions for single stars (red) and stars with planets (low-mass (green) and high-mass planets (blue)).}
                                \label{mgsi3sample}
                        \end{center}
                \end{figure}
                
                {We performed a K--S test to confirm these behaviours between the samples. As we can see in Table \ref{table:ks_co}, right column, we cannot reject the null hypothesis for the any planetary sample, so we have to assume that the samples come from the same distribution}.

                \section{C/O vs. Mg/Si: Implications on planet formation}
                
                We studied C/O  as a function of Mg/Si as a way to study the possible scenario for the formation of planets. In Fig. \ref{ratios1} we see how stars with and without planets are distributed in a C/O against Mg/Si plot. As can be seen, stars with planets, {both low-mass and high-mass,} are concentrated at C/O values of 0.5-0.6 and Mg/Si values of $\sim$ 1.0. The solar values of $\rm C/O_{\odot}=0.61$ and {$ \rm Mg/Si_{\odot}=1.07$}, derived from the adopted solar values for oxygen, carbon, magnesium, and silicon, are also represented.
                
                \begin{figure}
                        \centering
                        \begin{minipage}[c]{8.4cm}
                                \centering
                                \includegraphics[angle=180,width=1\linewidth]{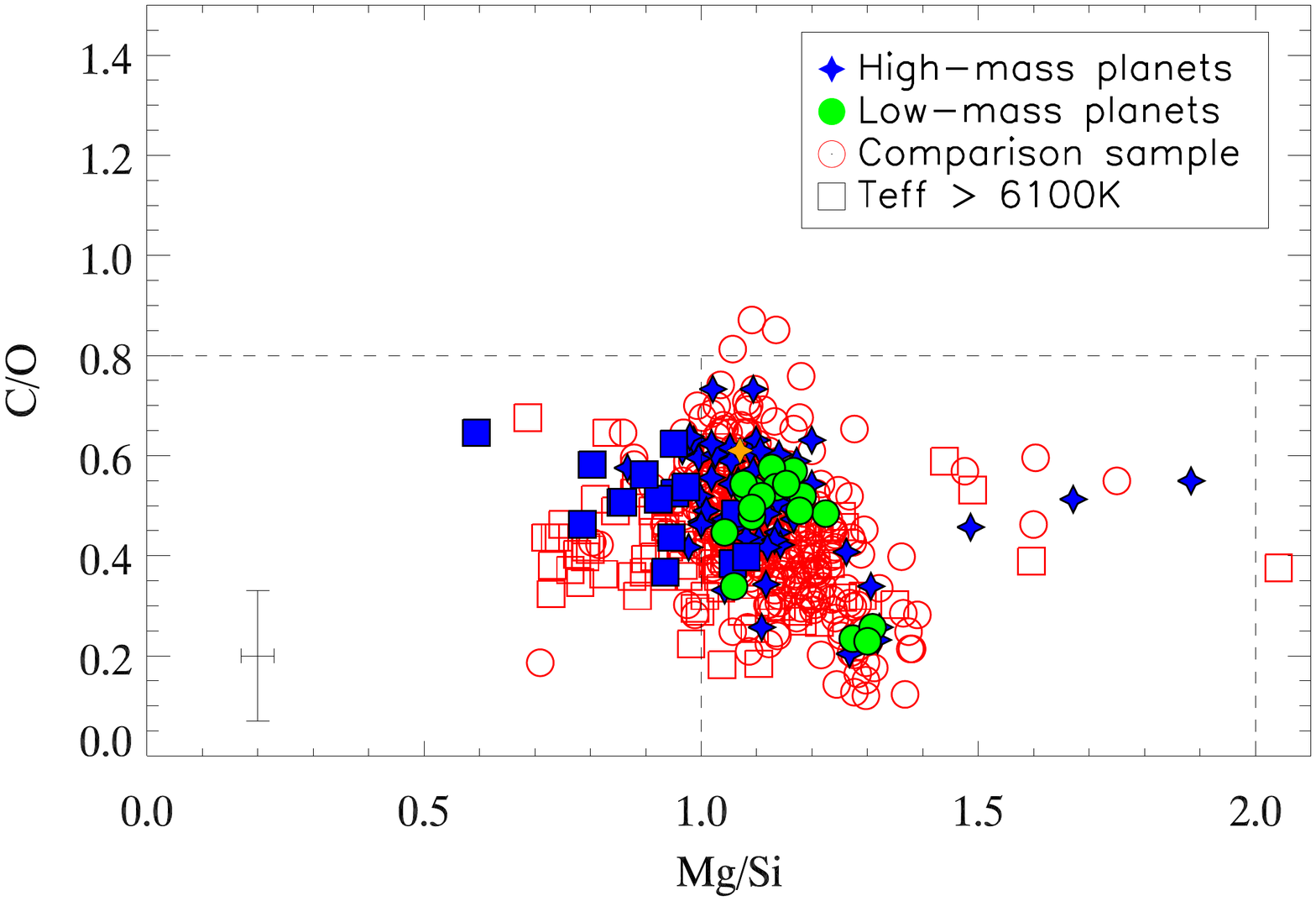}
                                \includegraphics[angle=180,width=1\linewidth]{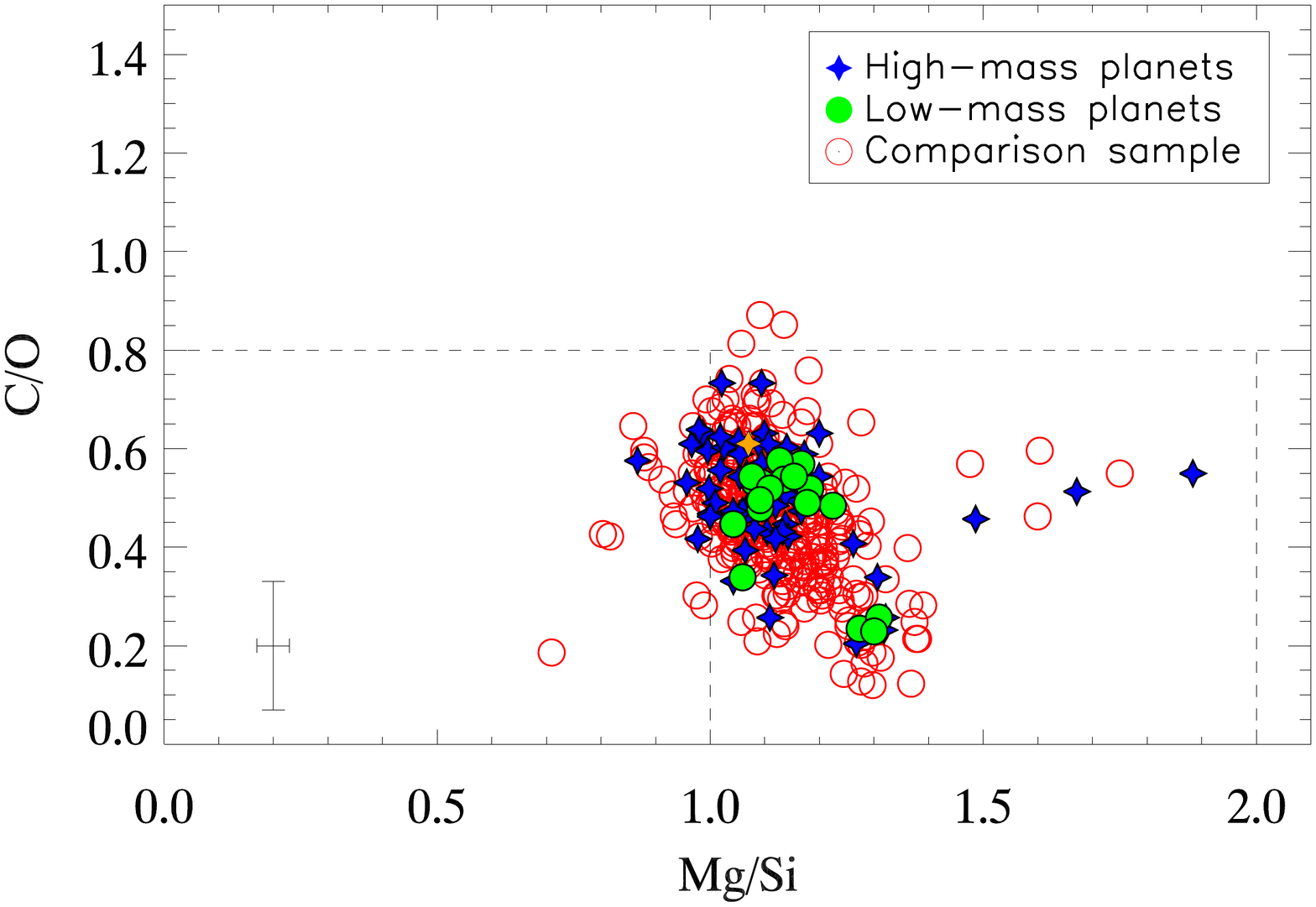}
                                \includegraphics[angle=180,width=1\linewidth]{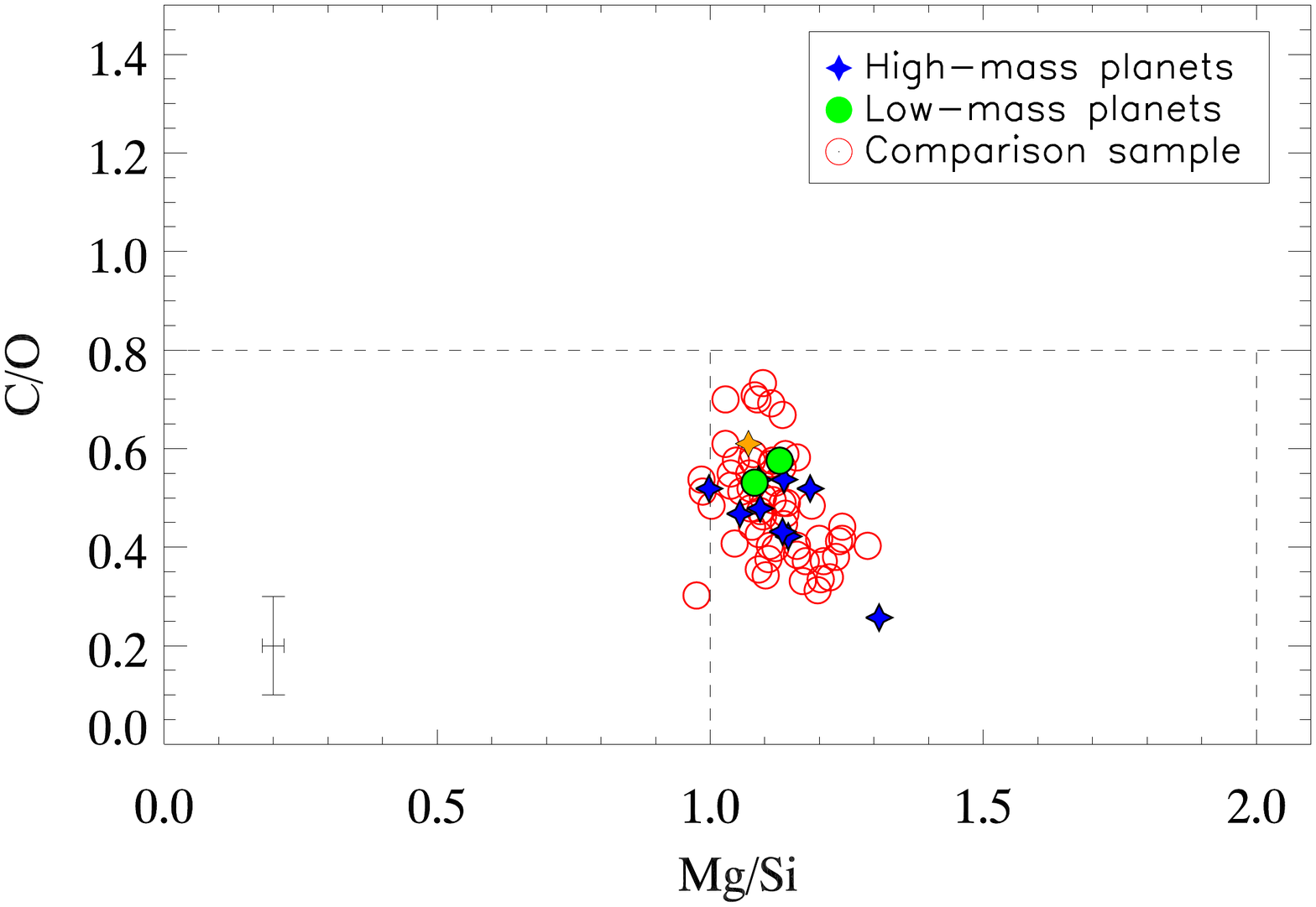}
                                \caption{Top panel: C/O vs.\ Mg/Si. Red circles refer to single stars (400 stars), and green dots refer to stars harbouring planets (99 stars). Middle panel: same as the top panel but only for stars with $\rm T_{eff} < 6100$K (312 single stars, 19 low-mass planet hosts and 65 high-mass planet hosts) . Bottom panel: same as top panel but for solar analogues,{ with $T_{\rm eff} = T_{\rm eff, \odot} \pm 300$ K, $\rm logg = \rm logg_{\odot} \pm 0.2$ dex and $\rm [Fe/H] = [Fe/H]_{\odot} \pm 0.2$ K (58 single stars, 2 low-mass planet hosts and 9 high-mass planet hosts).  In all panels, an orange star represents solar values.}}
                                \label{ratios1}
                        \end{minipage}
                \end{figure}
                
        {       As presented in Section 4, 100\% of the sample with low-mass companions have an Mg/Si value of between 1.0 and 2.0, while 85\% of the high-mass companion sample does, which means that Mg is equally distributed between pyroxene and olivine. We also find 15\% of high-mass planet hosts with Mg/Si values below 1.0 so Mg and Si will form mainly orthopyroxenes, whereas the remaining Si will take other forms, such as feldspars or olivine.  {No low-mass companions with Mg/Si values lower than 1.0 were found and no stars with Mg/Si values greater than 2.0 were found either. We highlight the fact that all stars with low-mass companions present Mg/Si $>$ 1.00, whereas the high-mass planetary sample can be found for a wide range of Mg/Si values, including Mg/Si < 1.00.}
                
                Regarding C/O, $100\%$ of stars with planets have C/O values lower than 0.8, meaning that Si will take solid form as $\rm SiO_{4}^{4-}$ and $\rm SiO_{2}$. {Only $\sim$15\% of our sample has C/O $<$ 0.4 (15 stars)}. The exact composition will be ruled by the magnesium-to-silicon ratio.

                {Recent models by \cite{carterbond,marboeuf14,thiabaud14,thiabaud15a,thiabaud15b} suggest that elemental abundance ratios may suffer great variations when studied in planet host stars and in planetary atmospheres. They proposed that elemental abundance ratios such as Mg/Si and Fe/Si would show similar values in both planet hosts and planetary atmospheres. As for C/O, they proposed large differences between the star and the planets.} They suggest that migration plays a key role when forming elemental abundance ratios in planetary atmospheres and the location of planet formation, as the total C/O ratios are governed by ices. \cite{oberg} studied the influence of different snowlines of oxygen and carbon-rich species. They suggest a common region between the $\rm H_{\rm 2}O$ and CO snowlines with giant planetary formation. These snowlines and the migration from the original birthplace can affect the abundance of volatiles, but not the abundance of refractory elements, such as Mg, Si, and Fe.  This assumption was confirmed for hot-Jupiter hosts by \cite{brewer16}, where they show that the average planet C/O is super-stellar, but their large uncertainties do not exclude the possibility of the 1:1 relation for the stellar and planetary C/O.
                
                {Stars with only low-mass planets are likely to be found in the 1 $<$ Mg/Si $<$ 1.5 regime, although mixed with stars with high-mass planets. As seen in Fig. \ref{ratios1}, there is a gap for stars with 1.1 $<$ Mg/Si $<$ 1.3  and C/O $<$ $\sim$0.5. 
                        As noted in Section 2, stars without planets are likely to be stars with undetected planets. Also, as presented in Section 4, there is a dependence in temperature for hotter stars. Since smaller planets will be harder to detect around these hotter stars, our results can be suffering from this effect. More knowledge about these stars and their plausible planetary companion is needed to constrain planetary formation and its relation with elemental abundance ratios more clearly.}

                \section{C/O, Mg/Si, and masses}
                
                We studied C/O and Mg/Si as a function of mass of planetary companion (the most massive planet in case of multiple planets). {In Fig. \ref{abundances} we can see a slight dependence on planetary mass, as the slopes are not null (-0.04 for Mg/Si and 0.04 for C/O), but the elemental abundance ratio errors do not allow us to make any further assumptions.}
                
                {As proposed by \citet{thiabaud15b}, Mg/Si in stars will give a direct information about the composition of the planet, as no differences are expected between them. Our stellar Mg/Si values can be translated as planetary Mg/Si \citep[see][]{santos15,dorn15}.
                        
                        Regarding C/O, and given the indirect relation between the star and the planet, our results are an estimation of the C/O that could be found in the planet. {For example, WASP-12b shows a high C/O, above 1 \citep{kreidberg}, while its host star shows C/O = 0.48 \citep{teske14}. \citet{brewer16} have proposed that planetary C/O will be super-stellar (while O/H could be sub-stellar) in hot-Jupiters.} 
                        
                        \begin{figure}[!ht]
                                \begin{center}
                                        \scalebox{1.1}[1.2]{
                                                \includegraphics[width=80mm, angle=180]{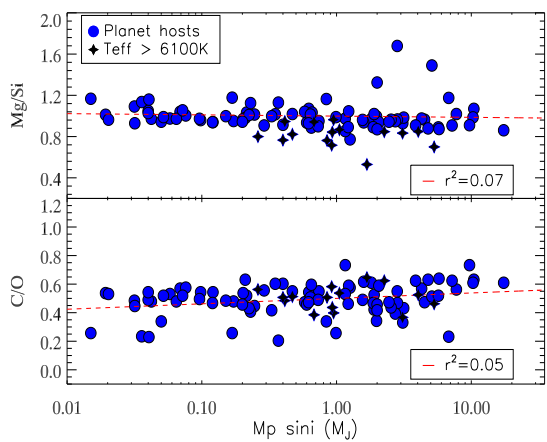}}
                                        \caption{{Mg/Si and C/O elemental abundance ratios as a function of the planetary mass. A correlation for the final sample (stars with $\rm T_{\rm eff}$ < 6100K) is provided for both sets, C/O and Mg/Si}.}
                                        \label{abundances}
                                \end{center}
                        \end{figure}

        \section{ Galactic trends in [C/O]}
        
        {Elemental abundance ratios can help us to understand how the planets are formed. In astronomy, solar ratios (e.g. [C/O]) are more frequently used, as opposed to absolute ratios (e.g. C/O). However, in order to investigate if the C/O ratios are affected by Galactic chemical evolution (GCE), \citep[as Mg/Si is, see][]{adi15}, we need to rely on solar ratios ([C/O]), therefore we present in this Section a study of [C/O] for our sample.
                
{Several works have already discussed the possible differences in individual elemental abundances and [X/Fe] abundances ratios between stars without detected planets and stars hosting low-mass and high-mass planets \citep{adi_over,adi12c,adi15}}. As presented in \citet{adi15} for a sample of 589 stars, the dependence of [Mg/Si] on metallicity, that is the Galactic evolution of [Mg/Si] ratio, may play an important role in the internal structure and composition of terrestrial planets. Overall, they concluded that stars formed at different times and places in the Galaxy have a different probability of forming low-mass planets, and the composition of the formed planets will also depend on the chemical composition of the environment in which they formed.

        { The dependence of [C/O] on metallicity, that is, the Galactic evolution of [C/O], may play an important role in the internal structure and composition of high-mass planets (as Mg/Si does for low-mass planets \cite[see][]{adi15}). Following this statement, we looked for possible relations between [C/O] and the masses of planetary companions.} We separated the planetary population into the same two groups as in the previous analysis (low-mass and high-mass planets). 

        In Fig. \ref{co_gce} we show the dependence of [C/O] on the metallicity for stars with and without detected planets.{ This figure shows that C/O has a clear dependance on [Fe/H]. To remove the trend of [C/O] with GCE we fitted all our data points (all stars with and without detected planets) with a quadratic dependence on metallicity and then subtracted the fit. Mean squared deviation for the fit is 0.06.}\footnote{[C/O]=-0.45*[Fe/H]$^{2}$+0.38*[Fe/H]-0.08}[C/O]$_{\rm corr}$ represents the [C/O] after substracting the fitted values. 
                \begin{figure}[!h]
                        \begin{center}
                                \scalebox{1.1}[1.1]{
                                        {\includegraphics[width=80mm, angle=0]{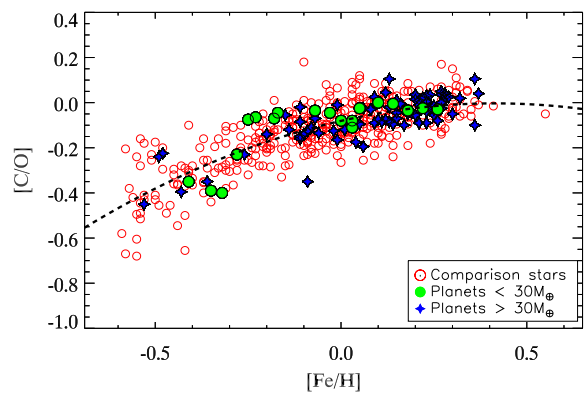}}}
                                \caption{[C/O] as a function of [Fe/H] for stars with and without detected companions. Trend line provides a {quadratic} fit to all the data points.}
                                \label{co_gce}
                        \end{center}
                \end{figure}
                
                \begin{figure}[!h]
                        \begin{center}
                                \scalebox{1.}[0.87]{
                                        {\includegraphics[width=80mm, angle=0]{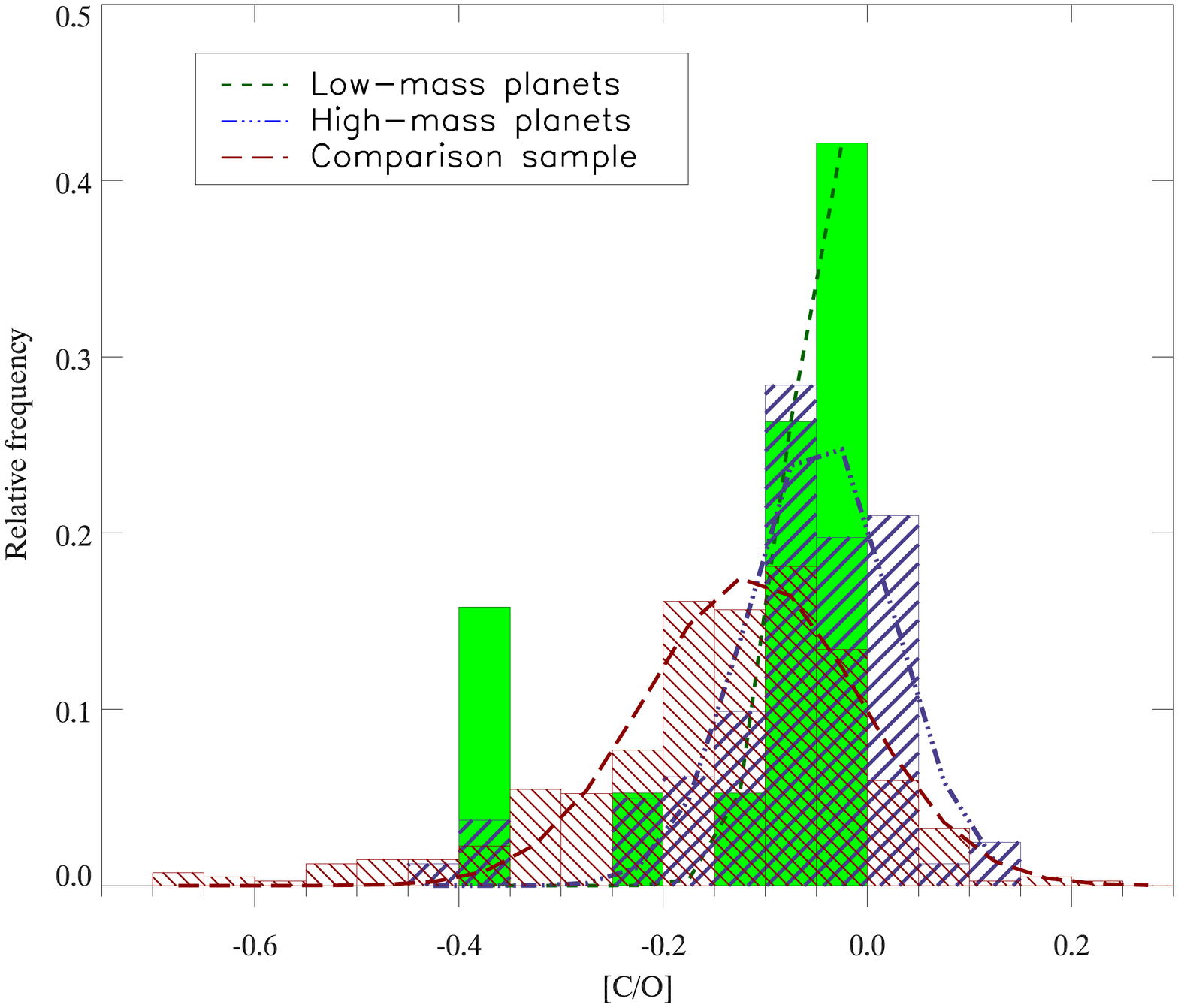}}}
                                
                                \scalebox{1.}[0.87]{
                                        {\includegraphics[width=80mm, angle=0]{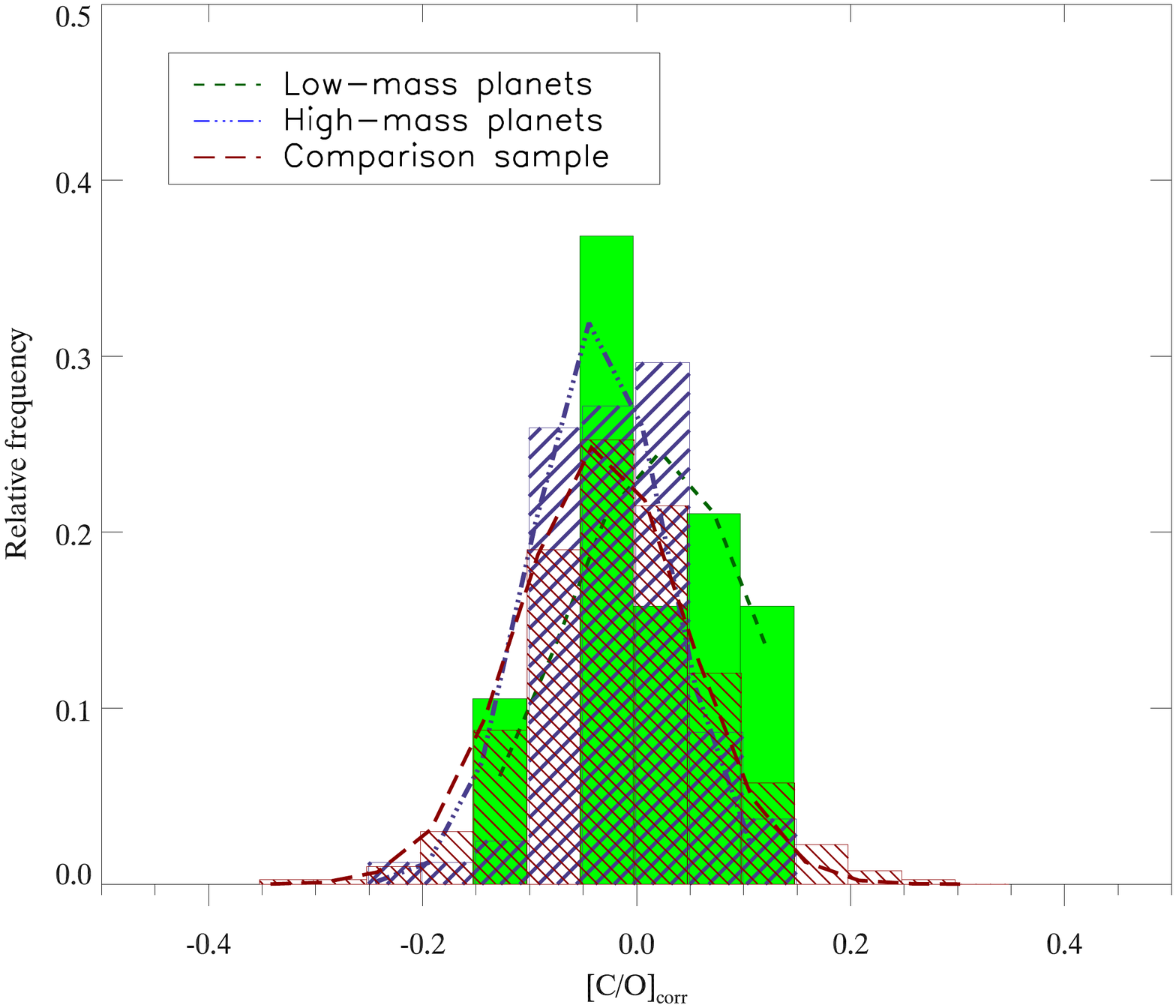}}}
                                \caption{[C/O] distributions before (upper panel) and after ($\rm [C/O]_{\rm corr}$, lower panel) correcting for the Galactic chemical evolution (GCE) effects.}
                                \label{histo_corr}
                        \end{center}
                \end{figure}
                
                Distributions of [C/O]  for all the presented sub-samples are shown in Fig. \ref{histo_corr}. A slight difference can be found between the centres of each Gaussian fitting, but they are within errors (see Table \ref{stats}). To obtain a clearer picture, we tested the sample using a two Kolmogorov-Smirnov (K--S) test (see Table \ref{table:ks_corr} and Fig. \ref{ks_co_ratio}). {The null hypothesis is rejected at level $\alpha$=0.1 if our results are higher than $D_{nn'}$=0.29 and $D_{nn'}$=0.15, for the low-mass and high-mass samples, respectively}. 
                {In the first case, [C/O], above the imposed threshold limit, the probabilities of similarity are 4\% and 4.09E-03\% for the low and high-mass sample, respectively, so we can assume that the samples do not come from the same distribution. In the second case, $\rm [C/O]_{corr}$, we cannot reject the null hypothesis, but, probabilities are 17\% and 38\%, respectively. Although we cannot assume any conclusion due to the result being below the threshold limit, we can see that probabilities of similarity rises significantly for the high-mass sample. The dependence of [C/O] on metallicity may play an important role in the internal structure and composition of high-mass planets (as Mg/Si does for low-mass planets \cite[see][]{adi15}) but more studies with more data are required to confirm or dismiss this assumption.}
                }

                \begin{figure}[!h]
                        \begin{center}
                                \scalebox{1.}[0.865]{
                                        {\includegraphics[width=80mm, angle=0]{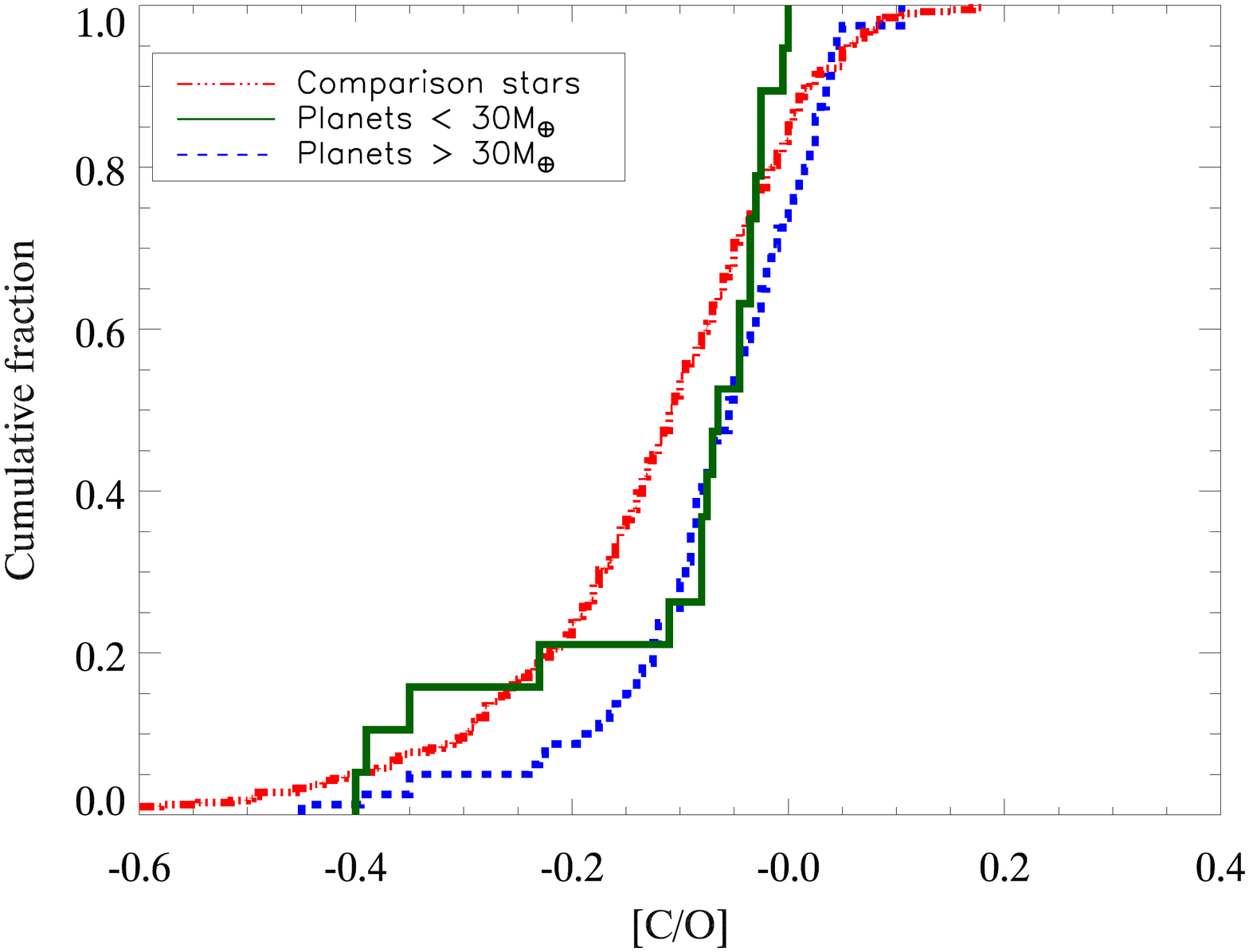}}}
                                
                                \scalebox{1.}[0.865]{
                                        {\includegraphics[width=80mm, angle=0]{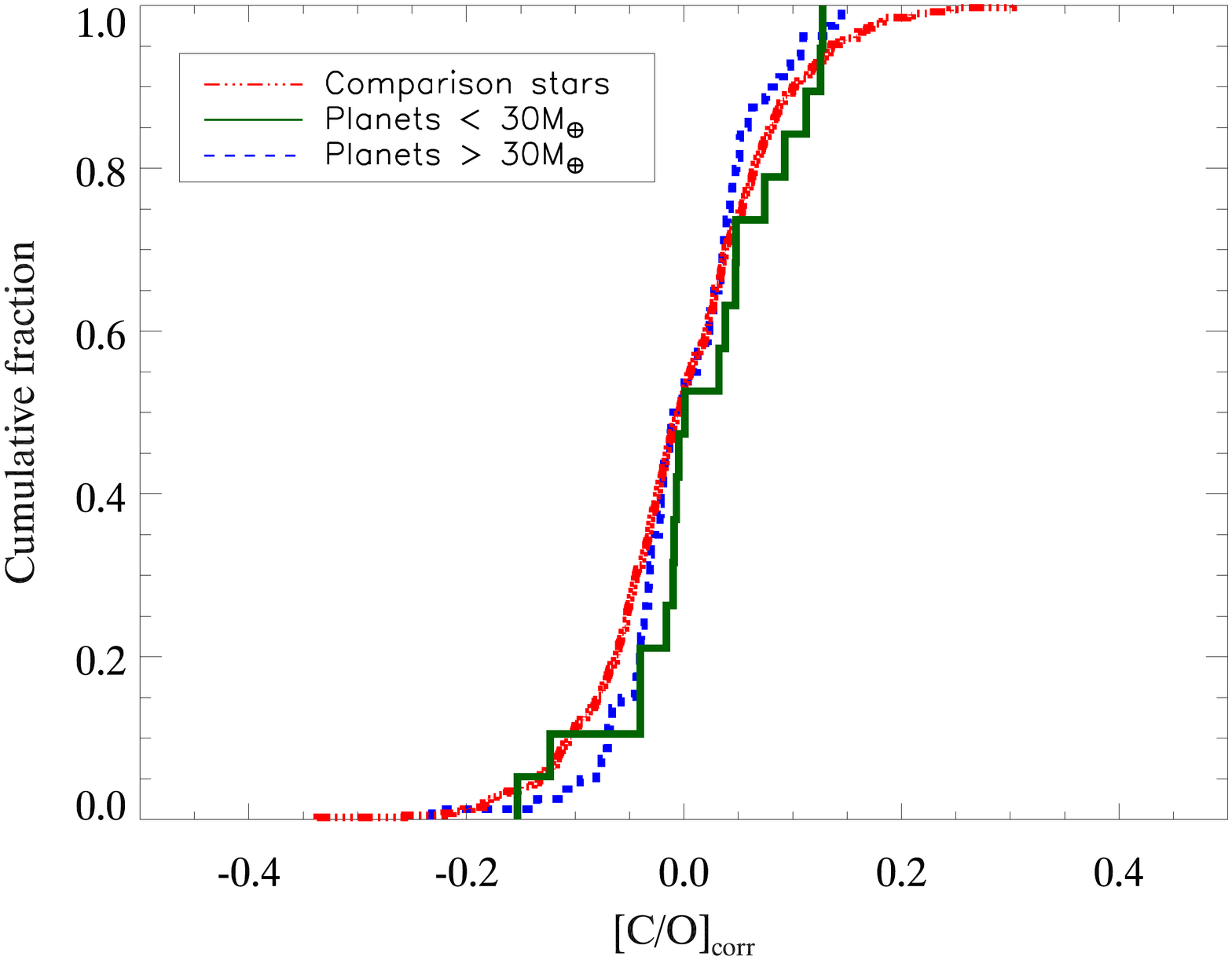}}}
                                \caption{Cumulative distributions for [C/O], before (upper panel) and after (lower panel)    the GCE correction. }
                                \label{ks_co_ratio}
                        \end{center}
                \end{figure}
                
                \begin{table}[!h]
                
                \caption{K--S test for GCE and non-GCE corrected samples.}
                
                \begin{center}
                        \scalebox{1.}[1.]{
                                \begin{tabular}{l c c c }
                                        \hline
                                        \hline
                                        
                                        Sample & [C/O] & $\rm [C/O]_{\rm corr}$ \\ 
                                        \midrule
                                        LMP - NP &  0.31 & 0.23 \\
                                        HMP - NP &   0.28 & 0.14 \\
                                        \bottomrule
                        \end{tabular}}
                        \label{table:ks_corr}
                \end{center}
        \end{table}

\section{Summary and conclusions}

We present a detailed study of the C/O and Mg/Si elemental abundance ratios for {499} solar-type stars observed with the HARPS high-resolution spectrograph. In our sample, 99 of {499} are planet-hosts and {400} stars have no known planetary companion. All the stars within our sample have effective temperatures between 5250 K and {6666} K, metallicities from $-0.59$ to $0.55$ dex and surface gravities from $3.81$ to $4.82$ dex.

We separated the planet population into two groups to test for possible relations between elemental abundance ratios and the masses of planetary companions: low-mass planets (LMP; with masses less than or equal
to 30 $M_{\oplus}$) and high-mass planets (HMP; with masses greater than 30 $M_{\oplus}$). All samples show the same distribution in a histogram, with al centre-fit similar within errors. Regarding the planetary sample, we cannot discern the high-mass from the low-mass sample.}

{We studied the probability of the samples being drawn from the same distribution by applying a K--S test to our samples. For the C/O case, our result suggest no similarity between the samples. For Mg/Si, we cannot reject the null hypothesis, so we have to assume that our samples come from the same distribution.  
Overall, 99\%  of our sample present C/O < 0.8, while all our confirmed planet host stars present C/O values below 0.8 (with peaks around 0.47), suggesting that the composition of the bulk of the planets should be $\rm SiO_{4}^{4-}$ and $\rm SiO_{2}$, {serving as seeds for}
Mg silicates. {$\sim$15\% of our sample has C/O < 0.4}. Regarding Mg/Si, and after separating our planetary sample in low-mass and high-mass companions, we found that none of our low-mass planet host sample present Mg/Si values lower than one, while 15\% of the high-mass sample does, suggesting a composition of pyroxene and feldspars for this sample. 

100\% of our low-mass sample present Mg/Si values between one and two, so an equal proportion of olivine and pyroxene is expected. 85\% of the high-mass sample present these values. These results agree (within the errors and taking into account different solar reference values) with recent studies of C/O and Mg/Si in the solar neighbourhood, such as \cite{brewer}, as they obtained peaks for distributions at $\sim$0.5 for C/O and $\sim$1.1 for Mg/Si. {Given these results, several different planet compositions could be found in planet hosts that meet these C/O and Mg/Si restrains.}

In the last few years, several models have suggested that C/O in planet host stars and planetary atmospheres is not the same, owing to migrations and snowlines, although more observations of atmospheres in transiting exoplanets are required to confirm these models {\citep{carterbond,oberg,marboeuf14,thiabaud14,thiabaud15a,thiabaud15b}}. 

These ratios give a suggestion of possible planetary abundances. As there is a direct relation between host star and planet abundances for Mg/Si,{ we can confirm the composition of the planets. Most of our planetary sample (100\% of the low-mass and 85\% of the high-mass sample) will have pyroxene and olivine as main components, while the other 15\% of high-mass companions will have a composition based on orthopyroxene and minerals such as feldspar or olivine. }

{For C/O, as the elements composing this ratio are sensitive to icelines, the observed planetary values can be up to three-four times the solar value, as recent observations suggest \citep{brewer16}. Planetary C/O values can be up to four times the proposed stellar values.}

{We studied [C/O] and its relation with metallicity \citep[as Mg/Si is affected, see][]{adi15}. We do not find a clear relation between high-mass planets and [C/O], but our results suggest that more data is needed to confirm this assumption. Our results for $\rm [C/O]_{corr}$ are below the threshold limit but suggesting a significant increase for the probability of similarity between the high-mass and the comparison star samples.} 
\begin{acknowledgements}
        We thank the reviewer for his/her thorough review and highly appreciate the comments and suggestions, which significantly contributed to improving the quality of the publication. J.I.G.H. acknowledges financial support from the Spanish Ministry of Economy and Competitiveness (MINECO) under 
        the 2013 Ram\'on y Cajal programme MINECO RYC-2013-14875, and the Spanish ministry project MINECO AYA2014-56359-P. V.Zh.A., E.D.M., N.C.S. and S.G.S. acknowledge the support from Funda\c{c}\~ao para a Ci\^encia e a Tecnologia (FCT) through national funds and from FEDER through COMPETE2020 by the following grants UID/FIS/04434/2013 \& POCI-01-0145-FEDER-007672, PTDC/FIS-AST/7073/2014 \& POCI-01-0145-FEDER-016880 and PTDC/FIS-AST/1526/2014 \& POCI-01-0145-FEDER-016886. V.Zh.A., E.D.M., N.C.S. and S.G.S. also acknowledge the support from FCT through Investigador FCT contracts IF/00650/2015, IF/00849/2015/, IF/00169/2012/CP0150/CT0002 and IF/00028/2014/CP1215/CT0002. This work has made use of the VALD database, operated at Uppsala University, the Institute of Astronomy RAS in Moscow, and the University of Vienna.

\end{acknowledgements}


\end{document}